\documentclass[11pt]{article}

\usepackage[final]{acl}

 \usepackage{microtype}

\usepackage[english,bidi=default]{babel} 
\usepackage{tgtermes}
\babelprovide[import]{hindi}
\usepackage{devanagari}
\babelprovide[import]{arabic}

\usepackage{courier}
\usepackage{subcaption}
\usepackage{soul}
\usepackage{xcolor}

\title{Privacy-preserving Prosody Representation Learning}

\author{Kevin Everson,  Mari Ostendorf \\
  Electrical \& Computer Engineering, University of Washington \\
  \texttt{\{kpever, ostendor\}@uw.edu} \\}

\begin{document}

\maketitle
\begin{abstract}
Speech representations that capture prosodic information can be useful for both understanding and generation. However, speaker characteristics are reflected in acoustic-prosodic features (e.g., pitch). To address privacy concerns from the leakage of identity information, we propose a new self-supervised approach to learning prosody representations that incorporates speaker disentanglement strategies. We evaluate our encoder on three tasks to probe representation capabilities, including pitch reconstruction and detection of different prosodic events. Our encoder outperforms raw prosody and HuBERT-base baselines, achieving strong speaker disentanglement without adverse impact on prosody-related downstream tasks.
\end{abstract}

\section{Introduction}

Prosody broadly refers to the aspects of speech which complement the lexical content, conveying intent-related or paralinguistic information. For example, local variations in pitch and energy, pausing, and duration lengthening can be used to communicate information focus, sarcasm, self-corrections, or the difference between a statement and a question. A wider pitch range and faster speaking rate can express excitement.  Representing prosodic information is essential for both speech generation and understanding, whether explicitly or implicitly in vector space models.  In this work, we introduce an explicit representation of prosody that aims to disentangle prosody from lexical content and speaker information, motivated by the need for more reliable understanding and generation of expressive speech and the desire for privacy-preserving speech processing systems.


Early computational work on prosody focused on acoustic cues such as fundamental frequency (F0), energy, and duration statistics, using speaker and phoneme normalization to remove speaker and lexical content information. Challenges with this approach are that automatic phonetic time alignment and F0 extraction algorithms are not very reliable, and energy is sensitive to recording conditions. More recently, researchers have explored self-supervised learning approaches, but often the objective of speaker disentanglement is lost.


Acoustic-prosodic cues are known to carry speaker information. 
When it is not removed, users are made vulnerable to serious privacy breaches, such as identity theft via deepfake generation. Concerns over these vulnerabilities are exacerbated by the proliferation of speech-based AI assistants and the ever-improving model capabilities. Thus, the disentanglement of speaker characteristics is essential for protecting user privacy in cases where identity information is not needed.

To this end, we train a prosody encoder by distilling information from acoustic-prosodic features via self-supervised learning, leveraging input features that disentangle lexical content and a learning framework for disentangling speaker information via target normalization and an adversarial loss function.
%
We evaluate the efficacy of our prosody representation on three tasks: a standard pitch reconstruction task, as well as prosodic prominence and phrase boundary detection.
The speaker identification task is used to assess our speaker disentanglement techniques. Our findings indicate that our encoder yields improved prosody modeling over baselines and that speaker information leakage can be effectively diminished without a negative impact on prosody modeling.

\section{Related Work}

While there have been a number of prosody representation models developed for specific tasks, e.g. speech synthesis, emotion recognion, and speech understanding tasks, this work focuses on the use of self-supervised learning with untranscribed speech to provide a general representation of prosody.
In addition, we are interested in frame-based representations, since they allow the flexibility of using different levels (e.g. phone, word) via pooling.

Standard acoustic modeling approaches used in speech recognition (e.g. HuBERT \citep{hsu_hubert_2021}, wav2vec 2.0 \citep{baevski_wav2vec_2020}) have proved to be useful for some prosody tasks \cite{lin_utility_2022}, but recently there have been a few efforts to more explicitly target prosodic characteristics of speech. 
%
%
In particular, our work builds on ProsodyBERT \cite{hu2023prosodybert} and PE-Wav2vec \cite{pe-wav2vec}.
As in ProsodyBERT, we use a HuBERT architecture with hidden units produced by acoustic-prosodic cues, and augment the standard masked prediction loss with a span boundary loss. However, like PE-Wav2vec, we use the estimated glottal waveform as input, rather than raw prosody features.
Different from both, we introduce speaker disentanglement strategies, including speaker-normalization of the prosody features used for masked prediction targets and an adversarial speaker loss function.
Other examples of speaker disentanglement in prosody representation learning include information bottlenecks \cite{qian_unsupervised_2021}, pitch-shifting the audio input \cite{weston_learning_2021}, and adversarial losses \cite{qu_disentanglement_2025}. 

Prosody representations have been evaluated in speech synthesis (or voice conversion), intent recognition (sentiment, sarcasm, persuasiveness), and paralinguistic tasks (emotion recognition, health diagnostics). Prosody benchmarks include SUPERB-prosody \cite{lin_utility_2022} and DAMMP \cite{weston_learning_2021}. Since we are interested in representation of linguistic (rather than paralinguistic) structure, we use prosodic event detection tasks, but also report on the pitch reconstruction task in SUPERB-prosody. 
Speaker disentanglement is most often evaluated with objective measures such as speaker identification or verification tasks \cite{lian_robust_2022, qian_autovc_2019}, or subjective assessments of source speaker characteristics present in generated speech \cite{deng_learning_2024, qian_unsupervised_2021}.
In this study, we use speaker identification.

\section{Prosody Encoder}

\begin{figure}[h]
    \centering
    \includegraphics[width=0.95\linewidth]{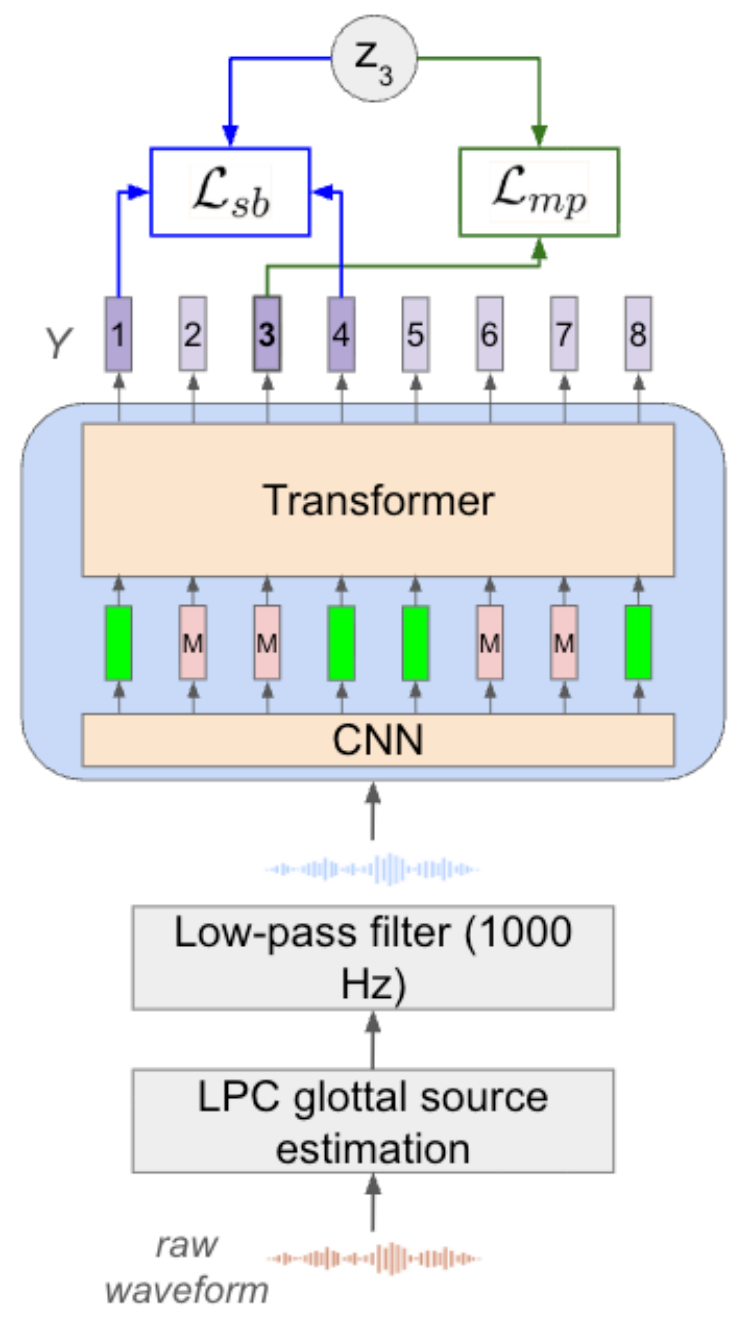}
    \caption{Prosody encoder pretraining, which uses a masked-prediction objective ${\cal{L}}_{mp}$ and a span-boundary objective ${\cal{L}}_{sb}$.}
    \label{fig:prosody_encoder}
\end{figure}

Our prosody encoder architecture, shown in Fig. \ref{fig:prosody_encoder}, is inspired by ProsodyBERT \cite{hu2023prosodybert}, which follows HuBERT's approach to self-supervised learning \cite{hsu_hubert_2021} but adds a span-based objective to encourage learning of suprasegmental characteristics. Key differences 
of our work include: i) use of the estimated glottal waveform as input, and ii) addition of an adversarial speaker identification loss term.


\subsection{Model details}


Our prosody encoder shares an architecture with HuBERT-base, consisting of a convolutional input module followed by a transformer.

\textbf{Input.} The model input is created using glottal source estimation, which is more robust and computationally efficient than pitch tracking, and captures voice quality information. We extract the glottal source by inverse-filtering with estimated LPC coefficients \cite{rabiner2010theory} (filter order 16, 25ms window, 10ms stride). In some frames with very low energy (non-speech regions), we find that the coefficients produced by LPC were unreliable, leading to input artifacts and training instability. To alleviate this issue, we simply return the raw waveform in frames with energy less than $10^{-4}$ rather than applying the inverse-filter. A $1 \text{ kHz}$ low-pass filter is then applied, yielding the final model input. In anecdotal listening experiments, we find that this low-pass filter reduces lexical information leakage. 

\textbf{Hidden units.} Similar to \citet{hu2023prosodybert}, hidden units $\mathcal{Z}=(\text{z}_1, ..., \text{z}_T)$ are produced via offline $k$-means clustering of acoustic-prosodic feature vectors $[\text{P}, \log \text{F}0, \Delta\log \text{F}0, \text{c}_1]$, where: 
\begin{itemize}
\setlength\itemsep{-0.3em}
    \item $\text{P}$ (periodicity) indicates the frame-level probability of voiced speech;
    \item $\log \text{F}0$, normalized by subtracting a speaker's mean log-pitch, where a speaker's mean log-pitch is computed as a weighted average over $\log \text{F}0$ using $\text{P}$ as weights;
    \item $\Delta \log \text{F}0$; and
    \item $\text{c}_1$, the first mel-frequency cepstral coefficient, representing the overall slope of the spectrum. 
\end{itemize}
Unlike \citet{hu2023prosodybert}, we normalize F0 features to support speaker disentanglement. $\text{c}_1$ is used in place of energy to reduce sensitivity to factors such as recording conditions.
Before clustering, we apply corpus-level $z$-normalization to each feature.

\subsection{Self-supervised learning objective}

The training loss function is the weighted sum:
$$
\mathcal{L} = \mathcal{L}_{mp} + \alpha_{sb} \mathcal{L}_{sb} + \alpha_{spk}^{adv} \mathcal{L}_{spk}^{adv},
$$
with hyperparameters $\alpha_{sb}$ and $\alpha_{spk}^{adv}$.

The first component, $\mathcal{L}_{mp}$, is the standard frame-level masked-prediction objective. For index $i$ in a masked span starting at index $j$ and ending at index $k$, a linear layer produces frame-level distributions over the $k$ hidden unit labels. The masked-prediction objective given by the cross-entropy:
$$
\mathcal{L}_{mp} = \log p_{mp}(\text{z}_i \:|\: \textbf{y}_i),
$$
where $\mathcal{Y}$ is the transformer output when randomly sampled spans of the convolutional encoder output are masked. We use the same span-level masking algorithm introduced in \citet{hsu_hubert_2021}.

To encourage learning of suprasegmental patterns, we add a span boundary objective \cite{joshi_spanbert_2020}. Specifically, we add a linear layer to predict the frame-level hidden unit at masked index $i$, given the \textit{nearest unmasked frames} on the left and right (at $j-1$ and $k+1$, respectively), as well as the distance (in frames) from $i$ to each. The span boundary objective is also given by the cross-entropy:
$$\mathcal{L}_{sb} = \log p_{sb}(\text{z}_i \:|\: \textbf{y}_{j-1},\:\textbf{y}_{k+1}, i-j+1, k+1-i).$$

Lastly, we include an adversarial speaker identification objective $\mathcal{L}_{spk}^{adv}$. The frame-level features $\textbf{y}_{i}$ are passed to a gradient-reversal layer, then projected via a linear prediction layer to produce a distribution over speaker labels:
$$
\mathcal{L}_{spk}^{adv} = \log p_{spk} (\text{spk} | \textbf{y}_{i}).
$$
Due to the gradient reversal layer, the linear prediction parameters are updated in the direction that minimizes $\mathcal{L}_{spk}^{adv}$, while the encoder parameters are updated in the opposite direction.

The prosody encoder is trained on the transcribed portion of the GigaSpeech corpus \cite{chen_gigaspeech_2021}, containing $11$K hours of English-language spontaneous and read speech from YouTube, podcasts, and audiobooks. Since speaker labels are not provided, we use pseudo-labels created by extracting utterance-level embeddings from a pretrained speaker encoder, then clustering with 1000 clusters per \cite{malinen_fixed-sized_2025}. Pitch and periodicity are extracted with \href{https://github.com/maxrmorrison/torchcrepe}{\texttt{torchcrepe}}.
We train using the \href{https://github.com/facebookresearch/fairseq}{\texttt{fairseq}} toolkit on four NVIDIA (A40 or L40) GPUs for $500$K steps, with a variable batch size averaging ${\small\sim} 30$ per GPU\footnote{\href{https://github.com/kpeverson/speaker_disentangled_prosody}{github.com/kpeverson/speaker\_disentangled\_prosody}}. The checkpoint with the lowest validation loss is frozen and used for downstream tasks.


\begin{figure*}[h]
    \centering



    {\fontfamily{lmtt}\selectfont \hl{such} super\hl{vi}sion | a\hl{ccor}ding to ash | is a \hl{sen}sible | \hl{cost} ef\hl{fec}tive al\hl{ter}native | to incarcer\hl{at}ion |}

    \caption{Prosody event example: ``|'' indicates a phrase boundary; highlighted text indicates prominent syllables.}
    \label{fig:bu_prosody_examples}
\end{figure*}

\section{Experiments and Results}
\label{sec:exp_setup}

We evaluate the encoder on a series of downstream tasks to assess its prosody modeling capacity and speaker disentanglement. The (frozen) encoder features are input to a task-specific model, which is fine-tuned on a single A40 or L40 GPU using the \href{https://github.com/s3prl/s3prl}{\texttt{s3prl}} toolkit\footnote{\href{https://github.com/kpeverson/s3prl\_tobi}{github.com/kpeverson/s3prl\_tobi}}. In line with our efforts to preserve privacy, we use only the final encoder output layer, in contrast to other SUPERB-based \cite{yang_superb_2021} setups, which use all intermediate layers. All reported results are on test splits, using the best checkpoint based on development splits.

\subsection{Evaluation models}

The proposed prosody encoder is compared to two baselines: the standard pretrained HuBERT-base model and the raw speaker-normalized prosody feature vectors ($[\text{P}, \log \text{F}0, \Delta\log \text{F}0, \text{c}_1]$).
For ablations, we train versions of our proposed prosody encoder without the adversarial speaker loss and without the speaker normalization of $\log \text{F}0$ when producing hidden units.

\subsection{Prosody prediction evaluations}

In addition to the standard pitch reconstruction task, we evaluate on two prosodic event detection tasks, chosen because the self-supervised learning approach was designed with the goal of characterizing linguistic (vs. paralinguistic) prosodic information.

\textbf{Pitch reconstruction.} 
We evaluate on the LibriTTS \cite{zen_libritts_2019} pitch reconstruction task using the setup and splits from SUPERB-prosody \cite{lin_utility_2022}. After extracting frame-level $\text{F}0$ offline using \href{https://bjbschmitt.github.io/AMFM_decompy/pYAAPT.html}{\texttt{pYAAPT}}, $\log \text{F}0$ is predicted over voiced frames from the prosody encoder features using a linear layer. Mean-squared error (MSE) is used as the training objective and reported metric.

Since average pitch varies with speaker and we aim to remove speaker information,
we also include a modified evaluation in which the predicted and ground-truth $\log \text{F}0$ contours are shifted to $0$-mean.

\textbf{Phrase boundary detection.} We introduce two tasks from the BU Radio Corpus \cite{ostendorf1995boston}, containing $11$ hours of read speech from seven professional radio announcers with word- and phoneme-level forced alignments, plus ToBI annonations \cite{beckman1994tobi} which include break index labels with word boundary timestamps. 
The corpus is partitioned such that ${\small\sim }80/10/10\%$ of each speaker's files are present in the training, development, and test splits, respectively.

\begin{table*}[h!]
\begin{tabular}{lcc||rr|rr|rr}
    & Speaker- &  & \multicolumn{4}{c|}{\textbf{ToBI detection tasks}} & \multicolumn{2}{c}{\textbf{Pitch recons.}}\\
    & norm. & \multirow{2}{*}{$\mathcal{L}^{adv}_{spk}$} & \multicolumn{2}{|c|}{Phrase boundary}  & \multicolumn{2}{c|}{Syl. prominence}  & Standard & $0$-mean \\
\textbf{Model}  & $\log \text{F}0$ &               & F1 ($\uparrow$) & acc. ($\uparrow$) & F1 ($\uparrow$)  & acc. ($\uparrow$)  &  MSE ($\downarrow$) & MSE ($\downarrow$) \\ \hhline{===||==|==|==}
\textit{most freq. class} & $-$ & $-$ & $\textit{0.00}$ & $\textit{0.87}$ & $\textit{0.00}$ & \textit{0.70} & $-$ & $-$ \\ \hline 
HuBERT-base & $-$ & \xmark & $0.79$ & $\textbf{0.95}$ & $0.74$ & $0.85$ & $0.056$ & $0.011$ \\ 
Raw prosody & \cmark & $-$ & $0.49$ & $0.88$ & $0.66$ & $0.83$ & $-$ & $-$ \\ \hline
\multirow{4}{*}{Ours} & \xmark & \xmark & $\textbf{0.82}$ & $\textbf{0.95}$ & $\textbf{0.86}$ & $\textbf{0.92}$ & $0.027$ & $0.012$ \\
    & \cmark & \xmark & $\textbf{0.82}$ & $\textbf{0.95}$ & $\textbf{0.86}$ & $\textbf{0.92}$ & $0.048$ & $0.012$ \\
    & \xmark & \cmark & $0.73$ & $0.93$ & $0.82$ & $0.89$ & $\textbf{0.024}$ & 0.012 \\
    & \cmark & \cmark & $\textbf{0.82}$ & $\textbf{0.95}$ & $\textbf{0.85}$ & $\textbf{0.91}$ & $\textbf{0.025}$ & $\textbf{0.008}$ \\
\end{tabular}
\caption{Prosody modeling \texttt{test} evaluation results.}
\label{tab:prosody_results}
\end{table*}
\begin{table}[]
\centering
\begin{tabular}{lcc||r}
& Speaker- &  & \\
& norm. & \multirow{2}{*}{$\mathcal{L}^{adv}_{spk}$} & Accuracy \\
\textbf{Model} & $\log \text{F}0$ & & ($\downarrow$) \\ \hhline{===||=}
 HuBERT-base & $-$ & \xmark & $0.64$ \\ \hline
\multicolumn{1}{l}{\multirow{4}{*}{Ours}} & \xmark & \xmark & $0.41$ \\
 & \cmark & \xmark & $0.42$ \\
 & \xmark & \cmark & $0.22$ \\
 & \cmark & \cmark & $\textbf{0.14}$ 
\end{tabular}
\caption{VoxCeleb1 \texttt{test} speaker identification results.}
\label{tab:spkr_id_results}
\end{table}

We detect full phrase boundaries (i.e. words with ``$4$'' break index labels, as indicated by ``|'' symbols in Fig. \ref{fig:bu_prosody_examples}) from prosody features within $\pm 100 \text{ ms}$ of each word boundary. Because changes in prosody can indicate phrase boundaries, we separately pool (via self-attention) the features before and after the timestamp, and concatenate the two resulting vectors before applying the classifier. This task is trained using cross-entropy loss, and evaluated using F1 score and accuracy.

\textbf{Syllable prominence detection.} The tone tier of the BU Radio Corpus ToBI annotations includes timestamps of accented syllables (containing ``*'' labels, indicated by highlighted text in Fig. \ref{fig:bu_prosody_examples}). To detect these, we pool the frame-level features from each syllable using self-attention and apply a linear classification layer to the syllable-level features. We use the same splits as in the phrase boundary detection task, and again use cross-entropy loss as the training objective and evaluate on F1 and accuracy.

\subsection{Speaker disentanglement evaluation}

To evaluate the amount of speaker information captured by our prosody modeling and the efficacy of our disentanglement methods, we also fine-tune on the VoxCeleb1 \cite{Nagrani2017} speaker identification task.

\subsection{Results and discussion}

Results on the prosody modeling evaluations are reported in Table \ref{tab:prosody_results}. Among the baseline systems, HuBERT-base achieves stronger results despite the lack of prosody-specific training strategies. The variants of our encoder outperform the baselines. The advantage of our system is greatest on the syllable prominence detection task, achieving a relative F1 improvement of $15\%$ over HuBERT-base.

Notably, the use of speaker-normalized $\log \text{F}0$ and the adversarial speaker objective together did not adversely affect downstream task performance; other than the variant with the adversarial speaker objective and without speaker-normalization of $\log \text{F}0$, the variants achieved similarly strong results on the ToBI detection tasks. At the same time, 
the variant with both speaker-disentanglement strategies performed the best in the $0$-mean pitch reconstruction setup, indicating the strongest modeling of local pitch dynamics.


The speaker identification (SID) results on the VoxCeleb1 test set are shown in Table \ref{tab:spkr_id_results}. We note that the accuracy of our HuBERT-base implementation is lower than the $0.81$ reported in \citet{yang_superb_2021}, since we only use the outputs of the final layer. 
%
Our prosody encoder variants indicate much less speaker information (lower SID accuracy) than HuBERT-base, and disentanglement strategies further diminish the speaker information. In terms of SID accuracy, the adversarial objective yields a 46\% relative reduction, and the two strategies together yield a 66\% relative reduction.


\section{Conclusion}

We propose a prosody encoder that takes the estimated glottal source as input, and is trained to predict hidden units produced from raw acoustic-prosodic features. Disentanglement techniques are employed to reduce the amount of speaker information present in the output features. Our findings indicate that prosody modeling is not negatively affected by these speaker disentanglement techniques, an encouraging result given the dual importance of capturing the intent-related information conveyed by prosody while protecting user privacy.

\section{Limitations}

We use pseudo-labels from utterance-level speaker embeddings, which places limitations on speaker-normalization of $\log \text{F}0$ and the adversarial speaker objective. Ground-truth labels would likely be more effective. \citet{chen_gigaspeech_2021} indicated plans to update the Gigaspeech corpus metadata with speaker information, but this has not yet been released.

The assessment of the representation learning approach is limited in a few respects. We focused primarily on local prosodic events, but  it may be useful to also evaluate on paralinguistic tasks. For comparison to prior work, the understanding tasks leverage hand transcriptions. Evaluation with automatically recognized transcripts would be more informative but would require a more sophisticated scoring algorithm. It would also be useful to evaluate on a speech generation task, which would provide an opportunity for subjective human assessments. Our model is not causal, which is needed for streaming generation scenarios, but translation of our approach to a causal framework is straightforward.

Comparisons to existing prosody models, namely ProsodyBERT \cite{hu2023prosodybert} and PE-Wav2Vec \cite{pe-wav2vec}, were limited due to lack of publicly available code. Our attempts to train similar models resulted in uncompetitive performance on downstream tasks.





\section{Ethical considerations}

The ethical concern of protecting privacy is an important consideration in the development of this model. Our approach was developed assuming relatively small amounts of data are available for a speaker.  It is possible that speaker recognition algorithms could be developed that are effective given larger amounts of data.


\section*{Acknowledgments}

This work was supported by the Intelligence Advanced Research Projects Activity (IARPA) via Department of Interior/Interior Business Center (DOI/IBC) contract number 140D0424C0070. The U.S.
Government is authorized to reproduce and distribute reprints for Governmental purposes notwithstanding any copyright annotation thereon. Disclaimer: The views and conclusions contained herein are those of the authors and should not be interpreted as necessarily representing the official policies or endorsements, either expressed or implied, of IARPA, DOI/IBC, or the U.S. Government.

\bibliography{custom}

@ARTICLE{pe-wav2vec,
  author={Liu, Zhao-Ci and Chen, Liping and Hu, Ya-Jun and Ling, Zhen-Hua and Pan, Jia},
  journal={IEEE/ACM Transactions on Audio, Speech, and Language Processing}, 
	title = {{PE-Wav2vec:} {A} {Prosody-Enhanced} {Speech} {Model} for {Self-Supervised} {Prosody} {Learning} in {TTS}},
  year={2024},
  volume={32},
  number={},
  pages={4199-4210},
  doi={10.1109/TASLP.2024.3449148}
}

@inproceedings{deng_learning_2024,
	title = {{Learning} {Expressive} {Disentangled} {Speech} {Representations} with {Soft} {Speech} {Units} and {Adversarial} {Style} {Augmentation}},
  author={Deng, Yimin and Wang, Jianzong and Zhang, Xulong and Cheng, Ning and Xiao, Jing},
  booktitle={2024 International Joint Conference on Neural Networks (IJCNN)},
  pages={1--7},
  year={2024},
  organization={IEEE}
}

@inproceedings{weston_learning_2021,
	title = {{Learning} {De-identified} {Representations} of {Prosody} from {Raw} {Audio}},
  author={Weston, Jack and Lenain, Raphael and Meepegama, Udeepa and Fristed, Emil},
  booktitle={International Conference on Machine Learning},
  pages={11134--11145},
  year={2021},
  organization={PMLR}
}

@inproceedings{qian_unsupervised_2021,
	title = {{Unsupervised} {Speech} {Decomposition} via {Triple} {Information} {Bottleneck}},
  author={Qian, Kaizhi and Zhang, Yang and Chang, Shiyu and Hasegawa-Johnson, Mark and Cox, David},
  booktitle={International Conference on Machine Learning},
  pages={7836--7846},
  year={2020},
  organization={PMLR}
}

@InProceedings{qian_autovc_2019,
	title = {{AutoVC:} {Zero-Shot} {Voice} {Style} {Transfer} with {Only} {Autoencoder} {Loss}},
  author =       {Qian, Kaizhi and Zhang, Yang and Chang, Shiyu and Yang, Xuesong and Hasegawa-Johnson, Mark},
  booktitle = 	 {Proceedings of the 36th International Conference on Machine Learning},
  pages = 	 {5210--5219},
  year = 	 {2019},
  editor = 	 {Chaudhuri, Kamalika and Salakhutdinov, Ruslan},
  volume = 	 {97},
  series = 	 {Proceedings of Machine Learning Research},
  month = 	 {09--15 Jun},
  publisher =    {PMLR},
  url = 	 {https://proceedings.mlr.press/v97/qian19c.html}
}

@inproceedings{lian_robust_2022,
	title = {{Robust} {Disentangled} {Variational} {Speech} {Representation} {Learning} for {Zero-shot} {Voice} {Conversion}},
  author={Lian, Jiachen and Zhang, Chunlei and Yu, Dong},
  booktitle={ICASSP 2022-2022 IEEE International Conference on Acoustics, Speech and Signal Processing (ICASSP)},
  pages={6572--6576},
  year={2022},
  organization={IEEE}
}

@misc{
hu2023prosodybert,
	title = {{ProsodyBERT:} {Self-Supervised} {Prosody} {Representation} for {Style-Controllable} {TTS}},
author={Yushi Hu and Chunlei Zhang and Jiatong Shi and Jiachen Lian and Mari Ostendorf and Dong Yu},
year={2023},
url={https://openreview.net/forum?id=7wk9PqiiW2D}
}

@inproceedings{lin_utility_2022,
	title = {{On} the {Utility} of {Self-supervised} Models for {Prosody-related} {Tasks}},
  author={Lin, Guan-Ting and Feng, Chi-Luen and Huang, Wei-Ping and Tseng, Yuan and Lin, Tzu-Han and Li, Chen-An and Lee, Hung-yi and Ward, Nigel G},
  booktitle={2022 IEEE Spoken Language Technology Workshop (SLT)},
  pages={1104--1111},
  year={2023},
  organization={IEEE}
}

@article{baevski_wav2vec_2020,
	title = {wav2vec 2.0: {A} framework for self-supervised learning of speech representations},
  author={Baevski, Alexei and Zhou, Yuhao and Mohamed, Abdelrahman and Auli, Michael},
  journal={Advances in neural information processing systems},
  volume={33},
  pages={12449--12460},
  year={2020}
}

@article{hsu_hubert_2021,
	title = {{HuBERT:} {Self-Supervised} {Speech} {Representation} Learning by {Masked} {Prediction} of {Hidden} {Units}},
  author={Hsu, Wei-Ning and Bolte, Benjamin and Tsai, Yao-Hung Hubert and Lakhotia, Kushal and Salakhutdinov, Ruslan and Mohamed, Abdelrahman},
  journal={IEEE/ACM transactions on audio, speech, and language processing},
  volume={29},
  pages={3451--3460},
  year={2021},
  publisher={IEEE}
}

@inproceedings{yang_superb_2021,
	title = {{SUPERB:} {Speech} {Processing} {Universal} {PERformance} {Benchmark}},
  author={Yang, Shu-wen and Chi, Po-Han and Chuang, Yung-Sung and Lai, Cheng-I Jeff and Lakhotia, Kushal and Lin, Yist Y and Liu, Andy T and Shi, Jiatong and Chang, Xuankai and Lin, Guan-Ting and others},
  booktitle={Proc. Interspeech 2021},
  pages={1194--1198},
  year={2021}
}

@article{ostendorf1995boston,
	title = {{The} {Boston} {University} {Radio} {News} {Corpus}},
  author={Ostendorf, Mari and Price, Patti J and Shattuck-Hufnagel, Stefanie},
  journal={Linguistic Data Consortium},
  pages={1--19},
  year={1995}
}

@book{rabiner2010theory,
	title = {{Theory} and {Applications} of {Digital} {Speech} {Processing}},
  author={Rabiner, Lawrence and Schafer, Ronald},
  year={2010},
  publisher={Prentice Hall Press}
}

@article{beckman1994tobi,
	title = {{The} {ToBI} {Annotation} {Conventions}},
  author={Beckman, Mary E and Hirschberg, Julia},
  journal={Ohio State University},
  url={https://www.cs.columbia.edu/~jjv/pubs/tobi_convent.pdf},
  year={1994}
}

@inproceedings{Nagrani2017, series={interspeech2017},
	title = {{VoxCeleb:} {A} {Large-Scale} {Speaker} {Identification} {Dataset}},
   url={http://dx.doi.org/10.21437/Interspeech.2017-950},
   DOI={10.21437/interspeech.2017-950},
   booktitle={Interspeech 2017},
   publisher={ISCA},
   author={Nagrani, Arsha and Chung, Joon Son and Zisserman, Andrew},
   year={2017},
   month=aug, collection={interspeech2017} }

@article{chen_gigaspeech_2021,
	title = {{GigaSpeech:} {An} {Evolving,} {Multi-Domain} {ASR} {Corpus} with 10,000 {Hours} of {Transcribed} {Audio}},
  author={Chen, Guoguo and Chai, Shuzhou and Wang, Guan-Bo and Du, Jiayu and Zhang, Wei-Qiang and Weng, Chao and Su, Dan and Povey, Daniel and Trmal, Jan and Zhang, Junbo and others},
  journal={Interspeech 2021},
  year={2021},
  publisher={ISCA}
}

@inproceedings{zen_libritts_2019,
	title = {{LibriTTS:} {A} {Corpus} {Derived} from {LibriSpeech} for {Text-to-Speech}},
  author={Zen, Heiga and Dang, Viet and Clark, Rob and Zhang, Yu and Weiss, Ron J and Jia, Ye and Chen, Zhifeng and Wu, Yonghui},
  booktitle={Proc. Interspeech 2019},
  pages={1526--1530},
  year={2019}
}

@misc{malinen_fixed-sized_2025,
	title = {{Fixed-sized} clusters {$k$-Means}},
	url = {http://arxiv.org/abs/2501.16113},
	doi = {10.48550/arXiv.2501.16113},
	urldate = {2025-10-14},
	publisher = {arXiv},
	author = {Malinen, Mikko I. and Fränti, Pasi},
	month = jan,
	year = {2025},
	note = {arXiv:2501.16113 [cs]},
}

@article{joshi_spanbert_2020,
	title = {{SpanBERT}: {Improving} {Pre}-training by {Representing} and {Predicting} {Spans}},
	volume = {8},
	shorttitle = {{SpanBERT}},
	url = {https://aclanthology.org/2020.tacl-1.5/},
	doi = {10.1162/tacl_a_00300},
	urldate = {2025-12-31},
	journal = {Transactions of the Association for Computational Linguistics},
	author = {Joshi, Mandar and Chen, Danqi and Liu, Yinhan and Weld, Daniel S. and Zettlemoyer, Luke and Levy, Omer},
	editor = {Johnson, Mark and Roark, Brian and Nenkova, Ani},
	year = {2020},
	note = {Place: Cambridge, MA
Publisher: MIT Press},
	pages = {64--77},
}

@article{qu_disentanglement_2025,
	title = {Disentanglement of {Prosody} {Representations} via {Diffusion} {Models} and {Scheduled} {Gradient} {Reversal}},
	volume = {36},
	issn = {2162-2388},
	url = {https://ieeexplore.ieee.org/document/10901998/},
	doi = {10.1109/TNNLS.2025.3534822},
	number = {8},
	urldate = {2026-01-04},
	journal = {IEEE Transactions on Neural Networks and Learning Systems},
	author = {Qu, Leyuan and Weber, Cornelius and Wang, Wei and Jin, Jia and Gao, Yingming and Li, Taihao and Wermter, Stefan},
	month = aug,
	year = {2025},
	pages = {15043--15054},
}




\end{document}